**Synergetic Behavior in 2D Layered Material/Complex Oxide Heterostructures**

*Kyeong Tae Kang, Jeongmin Park, Dongseok Suh\*, and Woo Seok Choi\**

K. T. Kang
Department of Physics, Sungkyunkwan University, Suwon 16419, Korea
J. Park
Department of Energy Sciences, Sungkyunkwan University, Suwon 16419, Korea
Prof. D. Suh
Department of Energy Sciences, Sungkyunkwan University, Suwon 16419, Korea
E-mail: energy.suh@skku.edu
Prof. W. S. Choi
Department of Physics, Sungkyunkwan University, Suwon 16419, Korea
E-mail: choiws@skku.edu

The marriage between a two-dimensional layered material (2DLM) and a complex transition metal oxide (TMO) results in a variety of physical and chemical phenomena that would not have been achieved in either material alone. Interesting recent discoveries in systems such as graphene/SrTiO$_3$, graphene/LaAlO$_3$/SrTiO$_3$, graphene/ferroelectric oxide, MoS$_2$/SrTiO$_3$, and FeSe/SrTiO$_3$ heterostructures include voltage scaling in field-effect transistors, charge state coupling across an interface, quantum conductance probing of the electrochemical activity, novel memory functions based on charge traps, and greatly enhanced superconductivity. In this progress report, we review various properties and functionalities appearing in numerous different 2DLM/TMO heterostructure systems. The results imply that the multidimensional heterostructure approach based on the disparate material systems leads to an entirely new platform for the study of condensed matter physics and materials science. The heterostructures are also highly relevant technologically, as each constituent material is a promising candidate for next-generation opto-electronic devices.

**1. Introduction**

Heterostructures of two distinct materials provide unprecedented opportunities for a wide range of intriguing and useful properties. Often, heterostructures do reveal functional properties that are not observed in their individual constituents. Interactions across the interface between the two materials can be considered from various viewpoints, e.g., structural, electronic, and magnetic coupling. These interactions alter the original lattice, charge, and spin degrees of freedom, resulting in unexpected yet technologically relevant physical behaviors. With the recent advancements in realizing and characterizing atomically thin layers of disparate materials and the technological advancements in miniaturizing the electronic devices, the concept of "interface is the device" is becoming increasingly relevant.[1]

Most of these heterostructures are realized using materials within the same structural families, such as, compound semiconductors, perovskite oxides, and more recently, van der Waals heterostructures.[1-4] These conventional heterostructures have their advantages in the epitaxial matching of lattices with minimized structural defects.[5-6] They are also easy to apply and allow control the homogeneous epitaxial strain. Nevertheless, heterostructures with structurally distinct layers can also be conceived. If the interface between the dissimilar layers can be well-defined, the heterostructure is equally feasible as the conventional heterostructure composed of the same structural families. This vastly expands the possibilities and potential of the heterostructures vastly, as the combination of the materials is now multidimensional.

Since the discovery of two-dimensional layered materials (2DLM), including graphene, numerous studies have focused on their behavior on different substrates. Ideally, freestanding 2DLM is theoretically plausible, but they are rather difficult to achieve experimentally. In particular, homogeneous control of strain is rather challenging for freestanding 2D layers. Therefore, the choice of the substrate materials for the 2DLMs is of prime importance in discovering, studying, and utilizing the novel properties of the heterostructures. While Si-based semiconductor materials or simple metals such as Cu or Au have predominantly been utilized as the substrates for 2DLMs,[7,8] design of more exciting properties is achieved by adopting functional materials. When a 2DLM is fabricated on top of a substrate material, a heterostructure composed of two distinct materials is naturally achieved.

Complex transition metal oxides (TMOs), or so-called functional oxides, foster a variety of exotic electrical and magnetic behaviors, including superconductivity, colossal magnetoresistance, 2D electron liquid, and multiferroicity.[9-13] The strongly correlated electronic nature originates from the strong polarizability of O ions in the inter-atomic scale, resulting in the versatile properties depending on the kind of transition metal element. With the recent advancement of atomic-scale epitaxy, an atomistic layer design of the physical properties of the TMOs is plausible, extending their potential with regard to "oxide electronics."

In terms of the heterostructures, oxides have long served as the "gate dielectric layer" within conventional semiconductor technology. Considerable effort is directed towards the development of ideal insulating oxides for Si-based technology. Therefore, insulating TMOs can be readily considered as natural candidates to form a heterostructure with 2DLMs. However, there are furthermore important advantages of using TMOs as a constituent material of a heterostructure. In the last two decades, a technique has been developed to achieve an atomically flat surface of perovskite oxide crystal, for the epitaxial growth of thin films.[14] Naturally, the thin films grown on top also have an atomically smooth surface and interface. This quality of TMOs ensures good homogeneous adhesion as well as minimized structural defects of the 2DLM with van der Waals nature.

In this progress report, we review various 2DLM/TMO heterostructures, which exhibit synergetic phenomena according to the formation of the heterointerfaces, as shown in **Figure 1**. While most previous studies focused on the physical behaviors of 2DLM alone, we also consider the role of the TMOs seriously. Indeed, the formation of the heterointerfaces, rather than the individual constituent material, is the cause of the unexpected behavior of the heterostructure. First, we discuss graphene/$SrTiO_3$ heterostructures as a prototypical hybrid system. Graphene and $SrTiO_3$ are the most important and famous model systems for 2DLM and perovskite TMO, respectively. Yet, the heterostructure composed of the two bears further surprises to be discovered. Second, the heterostructure of graphene and ferroelectric oxide is reviewed. Ferroelectricity is one of the major applications of perovskite oxides. For example, ferroelectric random access memory, ferroelectric gating in field-effect transistors, and ferroelectric tunnel junctions are actively pursued. The remnant polarization in the ferroelectric layer influences the charge carriers in graphene, altering the quantum transport behavior of graphene. Third, the monolayer FeSe/$SrTiO_3$ system is discussed. This system exhibits superconductivity with the highest critical temperature among the Fe-based superconductors, and a possible mechanism of the enhanced superconductivity is reviewed.

Finally, a more general and diverse 2DLM/TMO heterostructure is discussed, and a prediction for a future research direction in the field is presented. Although there are myriad combinations of 2DLM/TMO heterostructures, only a few well-studied materials have been considered thus far. Therefore, we predict tremendous possibilities for the emerging field of novel heterostructures.

## 2. Graphene/SrTiO$_3$ Heterostructures

Graphene is by far the most widely studied 2DLM. There are simple approaches to obtain a high-quality material, and it is chemically inert under ambient conditions.[15,16] It also shows intriguing quantum phenomena expected from a topological characteristics, including the quantum Hall effect.[17] On the other hand, SrTiO$_3$ is the most widely studied perovskite material to date. It is a prominent platform for oxide electronics as it is used as the substrate for many different versatile perovskite oxides with small lattice mismatch. It exhibits a large dielectric constant and a quantum paraelectric characteristic at low temperatures, which is closely related to the behavior of soft-mode phonon.[18] By introducing carriers via the creation of O vacancies or cation doping, the originally insulating material becomes metallic, and even superconducting, which is a popular research topic. The dynamics of O vacancies in SrTiO$_3$ is also of great interest, which is closely related to its resistance switching behavior.

As both graphene and SrTiO$_3$ are the prototypical materials of their respective material classes, it was natural for their heterostructure to be examined first. Research on the graphene/SrTiO$_3$ heterostructure began with the combination of graphene with bulk single-crystal SrTiO$_3$, which is commercially available. Later, a SrTiO$_3$ thin film was employed for better controllability. In this section, we review the studies on graphene/SrTiO$_3$ heterostructures.

### 2.1. Graphene/SrTiO$_3$ Bulk Heterostructures

To our knowledge, Akcöltekin *et al.* first demonstrated exfoliated graphene on a SrTiO$_3$ substrate in 2009.[19] The heterostructure was examined using an optical microscope to identify single-layer, bilayer, and few-layer graphene with good adhesion on the SrTiO$_3$ surface. In 2011, Bußmann *et al.* reported atomic force microscopy (AFM) and Kelvin probe force microscopy results for exfoliated graphene on a SrTiO$_3$ bulk crystal.[20] A systematic increase of the work function with an increasing number of graphene layers was observed, which was attributed to the substrate-induced electron doping in the graphene. Direct growth of graphene using deposition techniques such as chemical vapor deposition (CVD) on a SrTiO$_3$ substrate was also reported.[21] (**Figures 2a,b**) Such efforts can lead to large-area, highly uniform monolayer graphene synthesis on the SrTiO$_3$ substrate, facilitating research on this heterostructure.

Since the high dielectric permittivity is one of the most important properties of SrTiO$_3$, its influence on the transport properties of graphene was studied in detail by Couto *et al*.[22] A high dielectric constant is believed to provide efficient screening of charged impurities and enhance the electron mobility of a conducting material in general. This hypothesis has been widely adopted in perovskite oxides, such as SrTiO$_3$- and KTaO$_3$-based materials and heterostructures, which indeed show enhanced carrier mobility at a low temperature.[23-25] As the dominant scattering source in graphene is considered to be charged impurities in the substrate, the effective screening of the substrate was expected to enhance the carrier mobility in graphene. However, this has not actually been observed for graphene. The hypothesis was first proven invalid for graphene on different substrates, such as SiO$_2$, polymethylmethacrylate, bismuth strontium calcium copper oxide, and mica, for which the mobility does not show significant substrate (dielectric constant) dependence as shown in Figure 2c,d.[26] Graphene on a SrTiO$_3$ single-crystalline substrate also does not show a notable enhancement in the mobility at a low temperature, despite the large dielectric constant of SrTiO$_3$ (approximately 5,000 below 4 K).[22] Via bottom gating of the graphene-on-SrTiO$_3$-bulk (500 μm thick) device, Couto *et al.* discovered that the gate-field-dependent conductivity of graphene exhibits the same conventional V-shaped behavior with a charge-neutrality point as graphene on any other substrate (Figure 2e). The "robust" transport behavior of graphene, which is insensitive to the type of substrate, might imply that the charged impurities in the substrate are not the major source of scattering as shown in Figure 2f. Instead, a short-range resonant scattering due to impurities is suggested as the mobility-limiting scattering mechanism, which well-describes the conductivity away from the charge-neutrality region. Nevertheless, by applying a finite magnetic field, the effect of the SrTiO$_3$ substrate can be visualized, suggesting that the long-range electron-electron interaction near the $N = 0$ Landau level in the quantum Hall regime can be efficiently screened (Figure 2g).[22]

The transport behavior of graphene on a SrTiO$_3$ substrate was later theoretically explained by Sarma and Li.[27] The experimental charge carrier density dependent electric conductivity of graphene can be reproduced by introducing long-range Coulomb scattering ($\sigma(n) \sim n$) and short-range resonant scattering ($\sigma(n) \sim (n/n_i)\ln^2[(n/n_0)^{1/2}]$), where $n_i$ and $n_0$ are the carrier concentration of the resonant scattering defect and the normalization carrier density, respectively) for the regions near and away from the charge-neutrality point, respectively. The previous experiment disregarded the long-range Coulomb interaction supported by the fact that the behavior of the temperature dependence of the $N = 0$ longitudinal resistance of graphene on $SrTiO_3$ differs from that of graphene on other substrates. However, the theory suggests that there is intrinsic Coulomb scattering; this it is necessary to consider the existence of the low-density minimum conductivity.

The negligible influence of the large dielectric constant of $SrTiO_3$ and the importance of the intrinsic electron-electron interaction in graphene (at least for the case without any magnetic field) have further been studied via photoemission spectroscopy measurements. Ryu *et al.* observed the intrinsic correlated behavior of the graphene band structure, for graphene on a $SrTiO_3$ substrate.[28] Despite the use of CVD-grown graphene, the authors claim that the atomically flat $SrTiO_3$ provides an ideal platform for the study of the intrinsic electronic behavior of graphene, which is not possible using conventional $SiO_2$/Si substrate. The strongly correlated nature of the charge carriers in graphene, which is evidenced by the breakdown of the Wiedemnann-Franz law,[29] and the effect of the Dirac fluid, could be examined using the graphene-$SrTiO_3$ heterostructure.

As we have seen thus far, the detailed electronic properties of graphene on a $SrTiO_3$ substrate are rather distinctive for different approaches, i.e., transport measurements, spectroscopy, and theoretical calculations. The discrepancy may also arise from the different preparation techniques of graphene, i.e., exfoliation or direct deposition, and different measurement conditions, e.g., the temperature, magnetic field, or geometric shape of the graphene layer. Nevertheless, all studies on the graphene/$SrTiO_3$ heterostructure indicate that the high dielectric constant of $SrTiO_3$ does not influence the intrinsic electronic behavior of graphene. This conclusion may suggest that the dielectric "environment" of graphene is not changed even in the vicinity of the highly dielectric material $SrTiO_3$. One reason for this could be the van der Waals nature of graphene, which makes the distance between the graphene and $SrTiO_3$ as large as >3 Å.[30] On the other hand, it can also be argued that the "in-plane" transport behavior of graphene is not affected by the dielectric screening of the substrate along the "out-of-direction" of the heterostructure geometry. Even in perovskite oxide superlattices with strong covalent interactions between the layers (and consequently a small distance between the layers), the in-plane electron mobility of a layer is not as much enhanced as expected by the dielectric screening provided by the adjacent high dielectric permittivity layers.[23]

In addition to the dielectric screening, the SrTiO$_3$ crystal was also adopted as the substrate for the modification of the structural, electronic, and optical properties of graphene.[30,31] In particular, the termination layer of the (001) SrTiO$_3$ surface can be chosen to be either SrO or TiO$_2$, which differ in surface energy and chemistry.[32] Naturally, these differences lead to modifications in the crystal and electronic structures of graphene on top of SrTiO$_3$. It is more difficult to stabilize an SrO-terminated SrTiO$_3$ surface technically, whereas a TiO$_2$-terminated surface can be simply achieved via various etching techniques.[32] However, it is valid to investigate the discrepancy using theoretical calculations. Baran *et al.* used density functional theory calculations to show that graphene on TiO$_2$-terminated SrTiO$_3$ does not show a large change in the electronic structure, because of the weak van der Waals interaction.[30] On the other hand, graphene on SrO-terminated SrTiO$_3$ with a higher adsorption energy shows a larger interaction. The modification in the electronic structure is significant, and p-type electronic conductivity is expected in this case.

The high dielectric constant of SrTiO$_3$ plays a more important role for heterostructures based on thin film devices. While dielectric screening turns out to be not very efficient, the large dielectric constant is advantageous in scaling the gate voltage for a graphene field-effect transistor device, especially for the thin film case. Moreover, when epitaxially grown, the SrTiO$_3$ thin film offers an important platform for examining novel physical phenomena of the graphene/oxide heterostructure. The next subsection deals with the physics in graphene/SrTiO$_3$ thin film heterostructures.

### 2.2. Graphene/SrTiO$_3$ Thin Film Heterostructures

For the sake of comparison with the transport property of graphene/SrTiO$_3$ bulk, Sachs *et al.* investigated the transport behavior of graphene on an epitaxial SrTiO$_3$ thin film (250 nm) in 2014, as shown in **Figure 3a**.[33] To understand the mechanisms of Coulomb scattering, the transport behavior of graphene on various forms of SrTiO$_3$, e.g., an epitaxial thin film as well as bulk SrTiO$_3$ with different orientations, was examined. Using pulsed laser deposition, the epitaxial SrTiO$_3$ thin films were grown on the Nb-doped SrTiO$_3$ substrate, which can be adopted as a bottom electrode in gate measurements. Remarkably, graphene on both the bulk and the thin film SrTiO$_3$ exhibited hysteretic conductance but without the characteristic quantum Hall behavior. The hysteretic behavior is explained by the charge trapping or ferroelectric effects of the substrate. Sachs *et al*. suggested that the hysteresis does not come from the intrinsic property of graphene but rather from the $SrTiO_3$. Extrinsic defects - especially O vacancies that might serve as trapping sites - could be excluded because graphene/$SrTiO_3$ devices are post-annealed in O. Instead, the surface dipole of $SrTiO_3$ bulk and thin film was proposed, resulting in the ferroelectric-like contribution of $SrTiO_3$ to the hysteretic conductance (Figure 3b). The hysteretic transport behavior in the graphene/$SrTiO_3$ thin film (300 nm) was also reported by Saha *et al*., although the quantum Hall conductance of graphene was not observed again as shown in Figure 3c.[34] Likewise, the trapping sites and surface dipoles were attributed as the cause of the hysteresis in the conductance. The operating gate voltage ($V_G$) is reduced compared with that of conventional graphene on a Si-based substrate due to the high dielectric constant of $SrTiO_3$. However, it was rather difficult to quantitatively characterize the voltage scaling without the quantum Hall feature. In addition, a curious saturation of the carrier density at a certain value of electric field (~100 kV/cm) was observed (Figure 3d).

When the graphene and TMO thin film are both of high quality, the quantum Hall conductance arises. For example, the defects - such as the wrinkling of graphene, the rough surface of the TMO, or extrinsic molecules at the interface - distort the perfect two-dimensional (2D) electron system, and degrade the graphene transport performance. Fortunately, the TMO has the potential to be a well-suited substrate for 2DLMs owing to the high-level development for the pre-treatment, which results in an atomically flat surface with a well-determined chemical composition. For the case of the epitaxial thin film, the precise engineering of the stoichiometry during the fabrication process also plays a crucial role in characterizing the graphene transport.[30, 35] For example, owing to the multivalent nature of Ti within the $SrTiO_3$ thin film, only a tiny amount of defect sites can cause current leakage. Therefore, near-perfect stoichiometry of the $SrTiO_3$ thin film is necessary to examine the effect of the large electric field. Briefly, the observation of the quantum Hall transport ensures the high quality of both the graphene and the $SrTiO_3$ thin film underneath, which let us exclude the unintentional defect effects.

Park *et al.* reported the dramatic reduction of the operating voltage of a field-effect transistor comprising graphene on a high-quality $SrTiO_3$ thin film (300 nm) with the first observation of the quantum Hall conductance in such a heterostructure.[36] The epitaxial $SrTiO_3$ thin film was grown on the atomically flat Nb-doped SrTiO$_3$ single-crystal substrate by employing pulsed laser epitaxy. Figure 3e manifests that despite its thickness of hundreds of nanometers, the SrTiO$_3$ epitaxial thin film exhibits an atomically flat TiO$_2$-terminated surface with a step-terrace structure one unit cell in height. This ensures excellent adhesion of graphene, as was observed for the bulk SrTiO$_3$ substrates. Post-treatment is applied to the SrTiO$_3$ thin film to minimize the leakage current to <5 nA for the gating experiment. Without a magnetic field, the monolayer graphene demonstrates the usual V-shape conductance with the temperature-dependent shifts of the charge-neutrality point towards a negative $V_G$. While the high carrier density conductance shows negligible temperature dependence, peculiar temperature dependence is observed in the low carrier density regime, which is explained by long-range Coulomb scattering mechanisms, as previously discussed.[27] In the magnetotransport measurement at a low temperature, graphene manifests the quantum Hall conductance, at least up to 200 K (Figure 3f).[36] The quantum Hall plateaus are well-fitted to the conventional quantization phenomena of monolayer graphene, although the operating the range of $V_G$ is significantly reduced (~1/25) compared with that on the commercial silicon oxide substrates. Notably, the high dielectric constant and the small thickness of the SrTiO$_3$ thin film play essential roles in designing the miniaturized device, while maintaining the essential quantum conductance of graphene. For the SrTiO$_3$ thin film, the variation of the dielectric constant is directly examined with respect to the temperature, and it is recognized that the temperature-dependent shifts of the charge neutrality point follow the inverse of dielectric constant above 50 K. Conventionally, direct capacitance measurement of the dielectric constant of SrTiO$_3$ results in a combination of an extrinsic effect from the dead layer and distinctive effects of different electrodes.[37] As the quantized conductance is a measure of the charge environment in which the graphene sits, this approach provides an alternate method for the examination of the intrinsic dielectric behavior of the adjacent material.[36]

On top of the quantum Hall effect, the small thickness of the oxide thin film can easily host a large electric field, leading to the development of the redox-process-induced hysteretic conductance of the graphene device. It is rather well-established that the O vacancies can be generated/redistributed within the SrTiO$_3$ bulk/thin film via the electric-field-induced electrochemical process around highly active/metallic materials (**Figure 4a,b**).[38-40] Such a mechanism was intensively developed to explain the resistance switching behavior in TMOs in general, which might lead to neuromorphic device applications. As graphene can play the role of the active electrode, the large electric field along the thin film can result in the redox process, which introduces electron-devouring O vacancies. Even though the quality of the graphene/SrTiO$_3$ heterostructure is ideal, i.e., it readily shows quantum transport behavior, the intrinsic dynamics of O vacancies can lead to additional hysteresis. This hysteresis can be understood in terms of charge trapping by the introduction of the O vacancies. To characterize the underlying mechanism of the hysteresis and exclude other origins such as ferroelectricity, the analysis of the transport behavior is essential, e.g., the direction of the hysteresis and the $V_G$ dependence of the longitudinal current of the graphene.

As shown in Figure 4c, the hysteretic quantum Hall conductance in a graphene/SrTiO$_3$ thin film was first examined by Kang *et al*.[41] While the graphene/SrTiO$_3$ thin film (90 nm) shows typical linear conductance in the low field regime, a systematic development of the hysteresis appears as the electric field exceeds the threshold of 110 kV/cm. The saturation behavior of the source-drain current within the graphene continues unless $V_G$ is decreased, which is the essence of the hysteresis. When the positive $V_G$ is decreased, the current immediately decreases, which leads to a positive shift of the charge neutrality point. As $V_G$ moves towards the negative region, the current slope flattens but does not become thoroughly saturated. This results in the return of the charge neutrality point to the initial position. The SrTiO$_3$ thin film is stoichiometric within the experimental error, as confirmed by its insulating behavior and X-ray diffraction measurements. Additionally, the antihysteretic direction allows us to exclude the case of ferroelectricity. Therefore, the O vacancies intrinsically generated/annihilated by the electric field are considered to act as electron trapping sites. As the field exceeds the threshold value of ~110 kV/cm, the redox activity begins around the active metal, *e.g.,* graphene. The O vacancies start to accumulate at the interface between the graphene and the SrTiO$_3$ thin film. The field-induced trapping sites keep accommodating carriers that would have otherwise been injected into the graphene. This leads to the saturation and hysteresis in the conductance. The opposite reaction occurs when a sufficient negative field is applied, leading to the annihilation of O vacancies and the restoration of the charge neutrality point. Interestingly, quantum Hall conductance appears with the field-sensitive hysteresis in the presence of a magnetic field (Figure 4d). In addition, the width of the conductance plateau and the distance between the two charge neutrality points are proportional to the range of $V_G$, and the intrinsic changes in the dielectric properties of the SrTiO$_3$ thin film can be deduced from the precise analysis of the conductance, as follows:

$$d_{\text{OV}}\left(\frac{1}{\varepsilon_{\text{OV}}} - \frac{1}{\varepsilon_{\text{STO}}}\right) \propto \left(V_{\text{G-range}} - V_{\text{th}}\right). \tag{1}$$

This quantitative relationship between the dielectric properties and the thickness of O vacancy layer provides a route for understanding what happened within the TMOs as well as the promising potential of graphene as a tool to investigate the intrinsic properties in the TMO underneath (Figure 4e).

By inserting a perovskite $LaAlO_3$ thin film between the graphene and $SrTiO_3$, a reconfigurable multifunctional device can be obtained. The complex oxide heterostructure of the large bandgap insulators $LaAlO_3$ and $SrTiO_3$ has been enthusiastically studied owing to its intriguing metallic interface.[10, 42-43] Among the intriguing properties of the TMO interfaces, the metal-insulator transition was observed with respect to the thickness of the $LaAlO_3$ thin film, suggesting a polar catastrophe in this system.[42, 44] Just below the critical thickness of $LaAlO_3$ (4 u.c.), the conductive and insulating states can be locally switched via conductive atomic force microscope (c-AFM) lithography, introducing the nanoscale design of the electronic device.[45] Huang *et al*. first constructed the $LaAlO_3$/$SrTiO_3$-based field-effect transistor consisting of a graphene top gate by using c-AFM lithography.[46] Jnawali *et al*. investigated the magnetoconductance characteristics of a field-effect device comprising a graphene/$LaAlO_3$/$SrTiO_3$ heterostructure fabricated via c-AFM lithography.[47] The anomalous Hall effect and weak anti-localization effects reveal that the device exhibits the quantum Hall nature over a broad temperature range.

Similar to the graphene/thin-film hybrid field effect transistor that exhibits hysteretic conductance owing to its small thickness, the ferroelectric substrate can play a prominent role leading to controllable memristive properties of the device. The spontaneous polarization of a ferroelectric layer affects the transport behavior of graphene, such as the hysteretic conductance, allowing the design of next-generation memory. For some ferroelectric thin films, both the charge trapping effect and the ferroelectricity give rise to the peculiar transport behavior of graphene simultaneously. In the next section, the physical and technical issues for graphene/ferroelectric oxide bulk/thin film heterostructures are described.

## 3. Graphene/Ferroelectric Heterostructures

In the study of graphene, it is natural to consider effective methods for modulating the properties of graphene. Practically, researchers always place graphene on a substrate to manipulate it stably, because of its atomically thin feature. However, the substrate in contact with graphene always alters the intrinsic properties of the graphene.[48,49] For this reason, researchers have attempted to use an inert material such as silicon oxide or hexagonal BN as a substrate for graphene.[50,51] Alternatively, making a graphene device in the form of a suspended structure is preferred for revealing the intrinsic properties of graphene, for example, the fractional quantum Hall effect.[52-54] In contrast, one can switch the viewpoint from the study of intrinsic properties to the development of the controllability by actively employing the functionality in the supporting substrate. The combination of graphene and ferroelectric material originates from such an idea and was demonstrated in a heterostructure where graphene was in contact with a polymeric ferroelectric film such as poly(vinylidene fluoride-trifluoroethylene) (P(VDF-TrFE)) (**Figure 5a**).[55-59]

The system of graphene/ferroelectric-TMO heterostructures can be understood from two different viewpoints. One is the permanent modulation of the carrier density of graphene without external stimuli, such as an external electric field or chemical doping, which are conventionally employed in research on semiconducting materials. The other is the adoption of a charge-density-tunable material as an electrode for the ferroelectric capacitor. The carrier density modulation inside the electrode material in contact with ferroelectrics can possibly contain information regarding the ferroelectric TMO. Early studies on ferroelectric memory devices offer knowledge about the importance of the choice of metals for the electrode of ferroelectrics, which determines not only the change in the magnitude of polarization but also the reliability related to fatigue phenomena. This is analogous to the one-transistor-type ferroelectric memory modulating channel carrier density arising from the spontaneous ferroelectric polarization employed in the gate-dielectric layer.

**3.1. Graphene/Ferroelectric-Oxide Planar Device Configuration: Field-Effect Transistor**

$Pb(Zr_xTi_{1-x})O_3$ (PZT) is one of the representative thin-film ferroelectric TMOs, and graphene/PZT heterostructure devices have been fabricated using exfoliated or CVD-synthesized graphene on top of PZT thin films having an electrical contact on their back side. By the formation of source and drain contacts to graphene, its conductance can be evaluated on the surface of the PZT substrate. This device configuration of a field-effect transistor, using graphene as a channel, PZT as a gate dielectric, and a back-side contact as a gate electrode, helps the modulation of the ferroelectricity, which influences the charge conduction of graphene (Figure 5b). There are numerous reports using this standard device structure for the evaluation of the graphene/PZT system, indicating the clear modulation of the conductance by two different states of ferroelectric polarization (Figure 5c,d).[60-66] Owing to its well-known conduction characteristics, graphene can also be used as a sensor for monitoring the ferroelectric state of PZT.[67-69]

In 2010, Hong *et al.* reported a graphene/PZT heterostructure field-effect transistor.[61] They found that the conductance of graphene changed significantly because of the large hysteresis of the ferroelectric polarization depending on the gate bias. However, there was an anomaly in the hysteresis direction, which is opposite to the simple expectation. Ideally, there should be no variation of the conductance of graphene when the gate-bias is applied in a range larger than coercive voltages corresponding to saturated polarization states, and most of the changes in the conductance should occur in the ferroelectric switching region. However, the abnormal phenomenon called "antihysteresis" was also observed in other reports regardless of the type of graphene (exfoliated or CVD-synthesized). Baeumer *et al.* reported the more complicated antihysteresis-like behavior of a graphene/PZT transistor but argued that the carrier type modulation of graphene was caused by the ferroelectric polarization switching of a PZT capacitor.[64] From a device point of view, the permanent modulation of the conductance of graphene could be related to the memory device, which was evaluated in several studies, although the exact origin of the antihysteresis has not been clearly elucidated.

As there are various types of TMO thin films, many ferroelectric-TMOs have been evaluated with materials other than PZT as gate dielectrics. When $PbTiO_3$/$SrTiO_3$ superlattices served as a substrate for graphene, the tendency of the hysteresis direction varied depending on the temperature and the gate bias (Figure 5e,f).[70] On the bases of careful analysis of the surface state of the substrate, the origin of interfacial charge traps was classified into two different categories: an extrinsic factor related to the adsorbates from the environment and an intrinsic defect on the surface of the substrate. On the other hand, the modulation of the quantum Hall effect of graphene by the ferroelectric polarization was achieved in a graphene/$Ba_{1-x}Sr_xTiO_3$ thin film.[71] However, even in this case, there was a crossover between normal hysteresis and antihysteresis as the temperature varied.

In addition to ferroelectric TMO thin films, there are a few reports of the use of ferroelectric single crystals as a substrate for graphene. Jie *et al.* first reported the usage of a $[Pb(Mg_{1/3}Nb_{2/3})O_3]_{1-x}$-$[PbTiO_3]_x$ (PMN-PT) single-crystal substrate for a graphene transistor and revealed an interesting normal hysteresis direction for the CVD-synthesized graphene, unlike the case of the thin-film PZT substrate.[72] Furthermore, Park *et al.* fabricated a device having a structure of graphene/hexagonal-BN/PMN-PT and observed a transition between normal- and antihysteresis depending on the voltage-sweep ranges (Figure 5g).[73] In particular, the complicated electrical conductance variation in graphene, which is similar to that reported by Baeumer *et al.*, was analyzed on the basis of the ferroelectricity-involved charge trapping at the interface between hexagonal BN and PMNPT as shown in Figure 5h.[64] The role of interfacial charge traps is widely accepted as a crucial factor causing the antihysteresis behavior, as discussed previously. Indirect evidence for this phenomenon is the slow variation or large relaxation time of the conductance of graphene as a function of the time after polarization switching, which were reported for a graphene/PZT system as well as a graphene/PMN-PT system.[69, 74] The pyroelectric effect is also discussed in another report.

## 3.2. Graphene/Ferroelectric-Oxide Vertical Device Configuration: Other than Field-Effect Transistor

A vertical device having a graphene/ferroelectric-TMO/metal configuration was initially evaluated in the graphene/$BiFeO_3$/Pt heterostructure, which replaces the top indium-tin-oxide transparent conductor with graphene for a $BiFeO_3$-based ferroelectric photovoltaic cell.[75] Later, Yuan *et al.* developed this graphene/$BiFeO_3$ junction again in the shape of vertical tunneling devices, where the Au/$BiFeO_3$/graphene/$SiO_2$/Si-substrate heterostructure gave a gate-bias-tunable tunneling current.[76]

Lu *et al.* constructed a tunneling device structure of graphene/$BaTiO_3$/(La,Sr)$MnO_3$, representing the scheme of graphene/ferroelectric-TMO/metal according to the analogy of the ferroelectric tunnel junction, which has great potential for next-generation nonvolatile two-terminal memory.[77] They not only obtained a large on/off ratio depending on the polarization direction of $BiTiO_3$ but also identified the clear dependence of the device operation on the environmental conditions. Specifically, for molecules located between graphene and $BaTiO_3$, $H_2O$ gives unstable operation, but $NH_3$ molecules stabilize the tunnel junction device (**Figure 6a-g**). This information is important because the characteristics of interfacial charge traps between TMO and 2DLM are a critical issue for all 2DLM/TMO heterostructure devices. Using a similar vertical heterostructure device of graphene/$BaTiO_3$/(La,Sr)$MnO_3$, AFM was applied together with the technique of flexoelectric domain switching for nanodomain engineering.[78] The advantage of the atomically thin metallic feature of graphene uniquely enables this manipulation by flexoelectricity, which is useful for the study of graphene/ferroelectric-TMO heterostructures, in addition to the electrode-free device configuration.

The graphene/ferroelectric-TMO heterostructure can also be investigated in terms of the structural property variation revealed by Raman spectroscopy. From the piezoelectricity of the PMN-PT single-crystal substrate, the strain of graphene can be modulated by the PMN-PT substrate and monitored using a Raman signal.[79,80] Similarly, graphene on periodically poled LiNbO$_3$ (PPLN) has also been examined, where the different polarizations from opposite ferroelectric domains induce different amounts of charge transfer from the substrate to graphene, resulting in the gate-dependent peak shift of the Raman G-band (Figure 6h,i).[81]

## 4. FeSe/SrTiO$_3$ Heterostructures

A prominent example of a TMO (particularly SrTiO$_3$) that dramatically affects the electronic properties of the 2D layer is monolayer FeSe. The monolayer FeSe/SrTiO$_3$ heterostructure is the only system that exhibits a superconducting transition temperature ($T_c$) above 100 K,[82] other than the well-known cuprates. FeSe is the simplest Fe-based superconductor structurally and has a $T_c$ of 8 K in bulk.[83,84] However, when an FeSe monolayer is combined with a SrTiO$_3$ substrate, the $T_c$ rises as high as 109 K, which is well above the boiling point of liquid N (77 K). This astonishing enhancement of the superconducting $T_c$ was completely unexpected and underscores the importance of the "substrate" in characterizing the 2DLM above. More specifically, the interaction between the 2DLM and the TMO layer underneath can result in unprecedented phenomena, often improving the functional properties.

The phenomenon was first observed using in-situ scanning tunneling microscopy measurements by Wang *et al.* for an FeSe film grown on top of an SrTiO$_3$ substrate via molecular beam epitaxy (MBE).[85] A clear superconducting gap as large as ~20 meV was observed in scanning tunneling spectroscopy measurements, which was approximately one order of magnitude larger than that of bulk FeSe.[86] This leads to the preliminary expectation of a superconducting $T_c$ of ~80 K.

After its discovery in 2012, there have been several reports on the superconductivity of FeSe/SrTiO$_3$ heterostructures. Angle-resolved photoemission spectroscopy (ARPES) was extensively employed for examining the Fermi surface and the resultant origin of the superconductivity of the heterostructure.[87-91] The Fermi surface consists of only electron pockets near the zone corner, with a nearly isotropic superconducting gap (**Figure 7a**).[87, 91] The dependence on the strain[88] and carrier density[89, 92] led to the interpretation of the superconductivity in relation to the spin density wave and antiferromagnetic ground state, respectively. The role of the optical phonon mode of $SrTiO_3$ in the superconductivity of FeSe has also been investigated using ARPES, in which case the interfacial coupling to the charge carriers in FeSe corroborates the superconductivity.[90]

In addition to ARPES, direct transport and magnetic measurements were performed on the heterostructures to characterize the superconducting parameters,[82, 93, 94] such as the critical current density and critical magnetic field. In particular, Ge *et al.* employed in-situ four-probe transport measurement for the direct observation of the electrical resistivity drop, which revealed a $T_c$ of ~100 K, as shown in Figure 7b.[82] Furthermore, an electric field gating experiment combined with sample thickness control based on electrochemical etching indicated the important role of electrostatic doping.[95] Theoretical calculations have also been executed using various techniques, including two-stage renormalization group study, first-principles calculation, and the quantum Monte Carlo approach.[96-98]

Several different mechanisms have been suggested to explain the enhancement of the superconductivity in the $FeSe/SrTiO_3$ heterostructure. Naturally, the strong coupling of degrees of freedom across the interface between the 2DLM and TMO should be considered to explain the unprecedented $T_c$ enhancement. In particular, the strong electron-phonon coupling across the interface has been suggested, where the soft-mode phonon in $SrTiO_3$ can help the formation of Cooper pairs in the FeSe layer. The roles of the soft-mode phonon and ferroelectric instability in assisting the unconventional superconductivity in TMOs are beginning to be considered more seriously. In particular, in doped $SrTiO_3$, the polar nature was reported to possibly enhance the superconducting $T_c$.[99] It is rather uncommon to consider the role of phonons to be significant in identifying the superconductivity in Fe-based superconductors in general, as the unconventional superconducting system is known to be more related to the magnetic instability than to the lattice vibration. Nevertheless, experimental and theoretical findings indicate the importance of the soft-mode phonon in enhancing the superconductivity in the adjacent FeSe layer. In particular, it has been repeatedly suggested that the soft-mode phonon could indirectly assist the formation of Cooper pairs in the FeSe layer.

# 5. Other 2D Layered Materials on Complex Oxides

An important class of 2DLMs consists of transition metal dichalcogenides (TMDCs). Unlike graphene, they show a semiconducting property which can motivate the control of their electrical or optical properties, according to a basic textbook knowledge of semiconductor engineering. Molybdenum disulfide ($MoS_2$) is a representative n-type semiconductor among TMDCs and can be easily exfoliated from a bulk crystal in the form of few-layer flakes or synthesized in a monolayer shape. Conventional studies on the semiconducting properties of TMDCs and their usage as electrical and opto-electrical devices have been performed mostly on the surface of $SiO_2$ substrates.[100-104] However, the van der Waals contact between an atomically thin TMDC and functional TMOs can alter the properties of TMDCs owing to an interaction between them. Therefore, when TMOs are employed as a substrate for the TMDCs, the TMDC/TMO heterostructure implies an intriguing interaction across the interface.

From an experimental point of view, light is a useful method for testing this interaction, and the reduction of the charge transfer from the substrate of $LaAlO_3$ or $SrTiO_3$ to $MoS_2$ was achieved using photoluminescence variation for monolayer $MoS_2$.[105] Enhanced photoresponsivity was also reported in other materials, such as 2D black phosphorus on the surface of $SrTiO_3$ operating as a programmable photoconductive switch.[106] More direct evidence of the $MoS_2$-substrate interaction is observed for the heterostructure of $MoS_2$ on a periodically polarized lithium niobite (PPLN) substrate.[107] The different optical signal contrast matched well with the ferroelectric domain of the substrate resulting from the charge transfer between $MoS_2$ and PPLN. This result provides clear evidence for the ferroelectric control of $MoS_2$. These kinds of experimental approaches are based on the sensitive optical response of $MoS_2$ in the wavelength range of visible light.

As in the case of graphene, $MoS_2$ can also be considered as an electrode material. Furthermore, $MoS_2$ is controllable by the electric field when it is in contact with a ferroelectric TMO. With the possibility of tuning the electrical properties of $MoS_2$ in two different states according to the polarization of the ferroelectric TMO, this idea could be implemented in a ferroelectric tunnel junction device, for example, via the construction of $MoS_2/BaTiO_3/SrRuO_3$ heterostructures.[108] Even though there remains the issue of the interface between $MoS_2$ and $BaTiO_3$, the successful modulation of their electrical conduction was demonstrated experimentally (**Figure 8a-e**). A more common device platform using $MoS_2$/ferroelectric-TMO heterostructures was realized in ferroelectrically gated field-effect transistors with MoS$_2$ as a channel (Figure 8f). For a ferroelectric layer, PZT among TMOs and Al-doped hafnium oxide among high-dielectric-constant materials were tested owing to their compatibility with Si-based semiconducting fabrication processes.[109-111] Unlike the simple replacement of constituents with new materials in a conventional device, the synergetic effect could be observed in the MoS$_2$/PZT transistor device during its operation via optical modulation (Figure 8g,h).[112] With the strong light absorption by MoS$_2$, the adjacent ferroelectric thin film is affected, which induces the variation of the polarization strength (or switching) permanently. This can be interpreted as a method of optical control for ferroelectric devices.

## 6. Summary and Outlook

As we have seen so far, 2DLM/TMO heterostructures possess great potential both scientifically and technologically. The scientific aspect mainly arises from the combination of the quantum transport, semiconducting, low-dimensional, and superconducting behavior of the 2DLM with the dielectric, ferroelectric, magnetic, and electrocatalytic behavior of the TMO. In particular, the coupling across the van der Waals interface between the 2DLM and TMO is significantly stronger than expected, giving rise to a several non-trivial synergetic phenomena. The discovery of this behavior will lead to an exciting forefront of technological applications based on the heterostructure.

For wider study and utilization of 2DLM/TMO heterostructures, certain shortcomings must be overcome. In particular, several experimental limitations should be contemplated for the successful study of the heterostructure. First, the fabrication of the 2DLM/TMO heterostructure can be improved. Currently, high quality 2DLM/TMO heterostructure devices are mostly based on exfoliated 2DLM layers. In particular, characteristic physical properties such as the quantum Hall effect are only observed for such devices.[36, 41] On the other hand, the superconductivity in the FeSe/SrTiO$_3$ heterostructure is best observed for MBE-grown samples via in-situ measurement techniques such as ARPES and four-probe transport measurements.[82] Obviously, a delicate sample (in the case of FeSe) is degraded in air, possibly owing to molecular contamination on the sample surface. Such limitations might be lifted by studying the versatile growth mechanism of 2DLMs with particular emphasis on fabrication techniques such as CVD, which can also result in wide, homogeneous sample surfaces.[113] Second, the crystalline quality - especially the corresponding surface roughness of the TMO - strongly affects the performance of the heterostructure. In particular, the surface roughness determines the structural quality of the 2DLM layer, where an ideal 2D layer can be anticipated only on top of atomically flat TMO surfaces. Indeed, only those 2DLM layers with minimized structural defects exhibit the expected intrinsic physical behavior. While atomically flat surfaces are readily available for most of the commercially available TMO substrates, to fully utilize the versatile properties of the TMO, it should be fabricated as a thin film form. This further requires a careful growth procedure to ensure atomically smooth surfaces, somewhat limiting the possible material systems.[32] Finally, the van der Waals interface, i.e., the weak chemical bonding, implies that the lattice structures between the 2DLM and TMO will not be commensurate. This poses a consideration for a theoretical approach to the system, in particular for the first-principles calculation when considering the interaction between the layers, because the unit-cell of the heterostructure cannot be well defined.[30]

By overcoming the aforementioned limitations, we envisage that a whole new research field of heterostructures based on structurally distinctive materials can be introduced. While the 2DLM/TMO heterostructure is attractive, as the 2DLM has a van der Waals bonding character and the TMO has a covalent (or ionic) bonding character, the formation of the interface will alter the chemistry as well as the electronic structure. We can further imagine heterostructures composed of other constituent materials, according to our understanding of this unconventional heterostructure. On the other hand, heterostructures with the opposite sequence can also be envisaged. That is, the TMO/2DLM heterostructure can be realized, for example, by using the graphene layer as a substrate for the TMO thin film. This approach has resulted in an $SrTiO_3$/graphene heterostructure,[114] with a bonding nature that cannot be conventionally categorized. Also the recent successful fabrication of transferrable 2D TMO layers will also enrich the possible device configurations, and thus, possibility of observing and studying exotic physical behaviors.[115,116] In conclusion, our understanding of 2DLM/TMO heterostructures and the proliferation of future studies in this field of research will result in an unexpected technological impact on the opto-electronic and magnetic devices.

**Acknowledgements**
K.T.K and J. P contributed equally to this work. This work was supported by Basic Science Research Programs through the National Research Foundation of Korea (NRF-2017R1A2B4011083) and by the Ministry of Trade, Industry and Energy and Defense Acquisition Program Administration through Civil Military Technology program (16-CM-MA-14). It was also supported by the Institute for Basic Science (IBS-R011-D1).

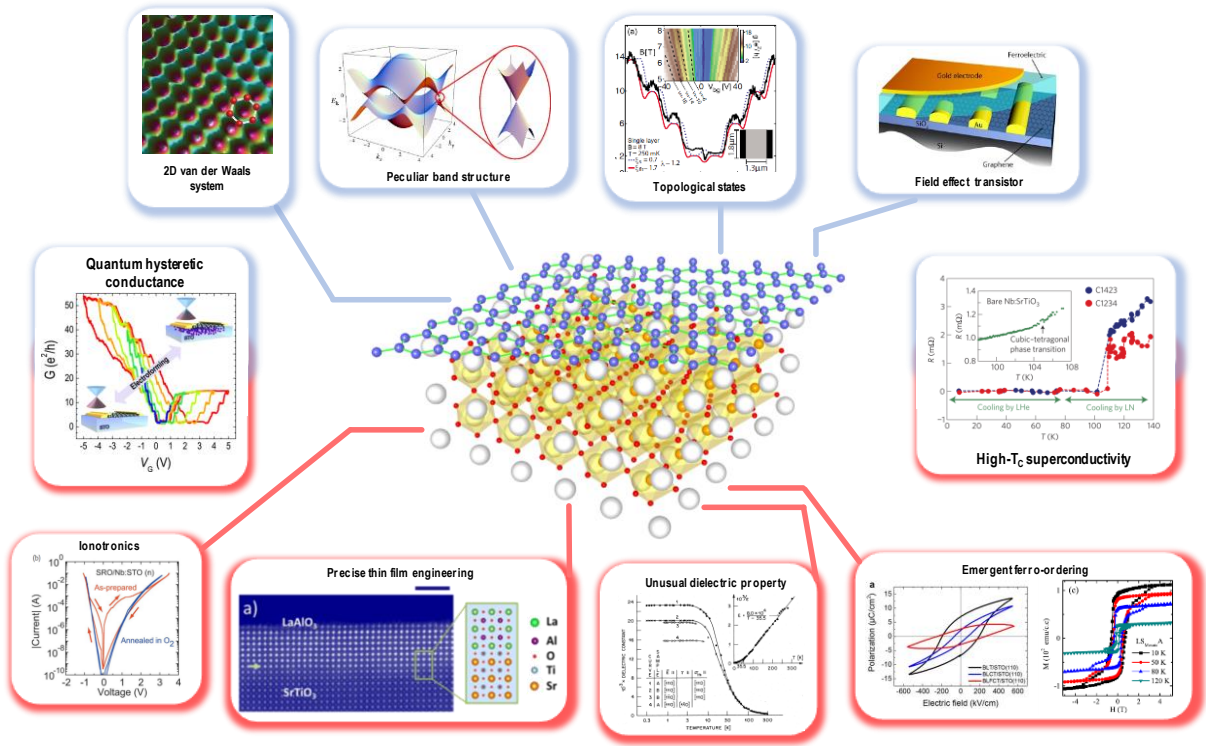

**Figure 1.**
Synergetic effect on heterostructures of two-dimensional layered material (2DLM) and complex transition metal oxide (TMO) material.
Reproduced with permission.[12] Copyright 2014, Nature Publishing Group. Reproduced with permission.[13] Copyright 2016, Nature Publishing Group. Reproduced with permission.[15] Copyright 2009, American Physical Society. Reproduced with permission.[16] Copyright 2011, American Institute of Physics. Reproduced with permission.[17] Copyright 2009, American Physical Society. Reproduced with permission.[39] Copyright 2008, Elsevier Ltd. Reproduced with permission.[41] Copyright 2017, Wiley-VCH. Reproduced with permission.[43] Copyright 2014, American Physical Society. Reproduced with permission.[55] Copyright 2009, American Institute of Physics. Reproduced with permission.[82] Copyright 2014, Nature Publishing Group.

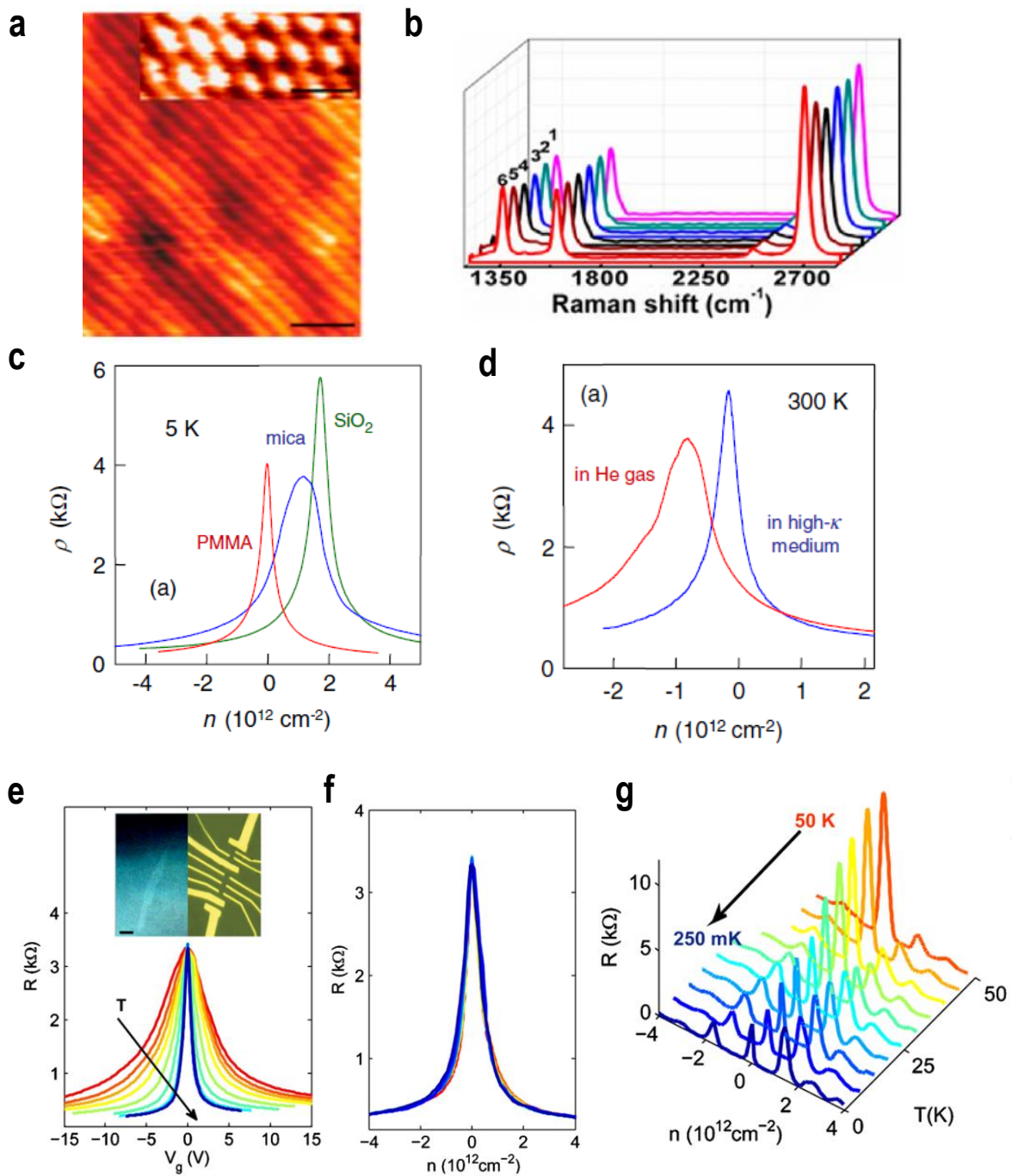

**Figure 2.**
(a) STM image of a graphene of hexagonal lattice on an STO substrate. The scale bar indicates 1 μm. (b) The Raman spectra which are collected from the random 130 points over a 3×3 mm$^2$ of the large area graphene grown by CVD, suggesting the uniformity of produced film. Reproduced with permission.[21] Copyright 2014, American Chemical Society. (c-d) Transport behavior of graphene in the different dielectric environments. To figure out the dielectric screening effects on graphene, high-k environment of glycerol ($k \sim 42$) is employed, but the remarkable enhancement of electron mobility is not observed even at low temperature. Reproduced with permission.[26] Copyright 2009, American Physical Society. (e-f) Robust transport behavior of graphene on the SrTiO$_3$ bulk insensitive to the substrate which suggests a short-range scattering mechanism. (g) Longitudinal square resistance measurement at 15 T to examine quantum Hall conductance. Herein, the stark decrease of the peak height has been observed with respect to the temperature decrease. Reproduced with permission.[22] Copyright 2011, American Physical Society.

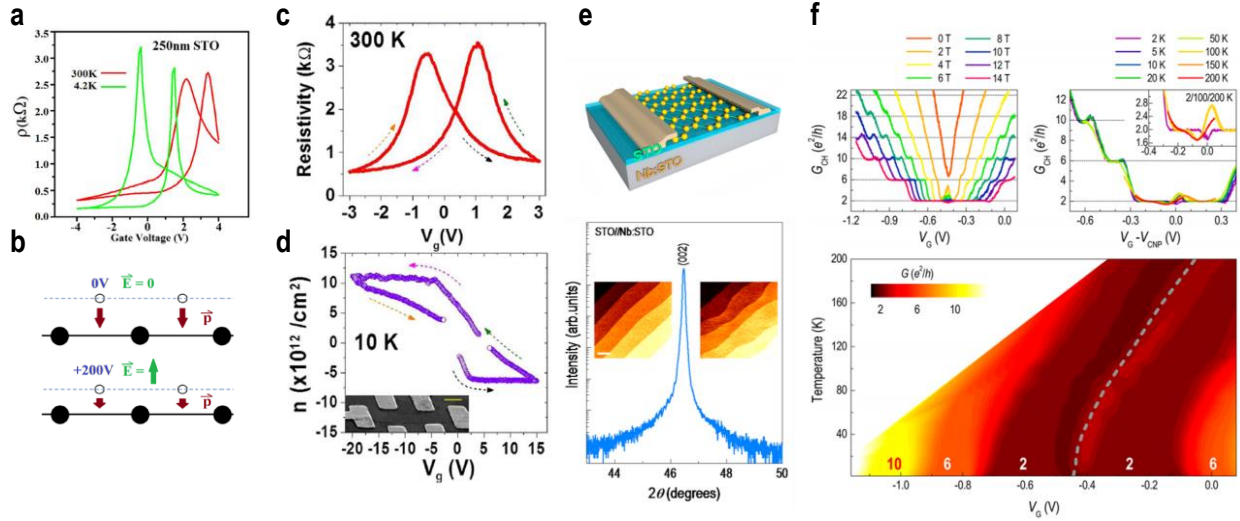

**Figure 3.**
(a-b) Hysteretic conductance behavior of graphene on SrTiO$_3$ thin film (250 nm). The scenario of the surface dipole of the SrTiO$_3$ thin film has been employed to understand the ferroelectric-like substrate phenomena. Reproduced with permission.[33] Copyright 2014, Nature Publishing Group. (c-d) Hysteretic conductance behavior of graphene on SrTiO$_3$ thin film (300 nm) with a notable reduction of operating voltage. Although the hysteresis can be understood in terms of charge trapping, a curious saturation of the carrier density is observed at ~100 kV/cm. Reproduced with permission.[34] Copyright 2014, Nature Publishing Group. (e) Schematic of monolayer graphene on SrTiO$_3$ thin film device. To exclude extrinsic effects, the crystalline and atomically flat substrate is prepared. (f) Channel magnetotransport quantization as a function of gate voltage at 2 K under various magnetic field and under 14 T magnetic field at various temperatures. The temperature-dependent shift of charge neutrality point (gray line) arises which is attributed to the temperature-dependent change of dielectric nature of substrate underneath. Reproduced with permission.[36] Copyright 2016, American Chemical Society.

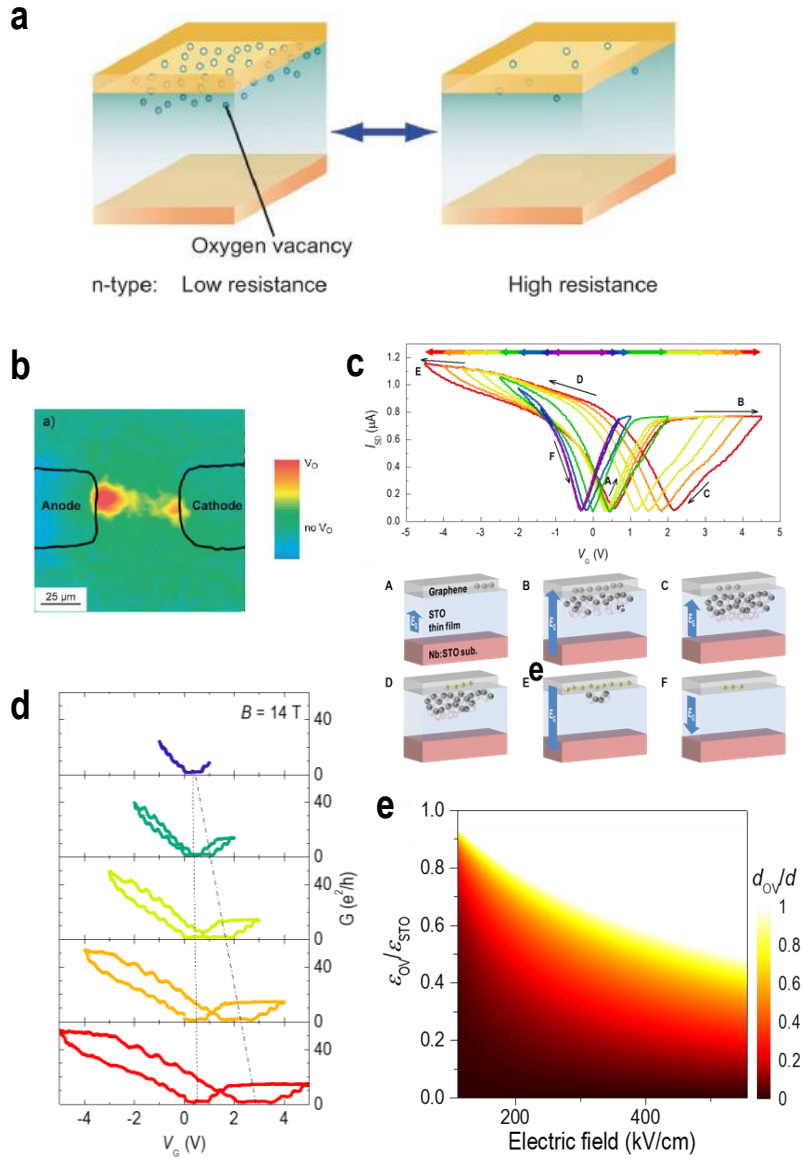

**Figure 4.**
(a) Oxygen vacancies can be created/annihilated by electric field application, which leads to the resistive state switching. Reproduced with permission.[39] Copyright 2008, Elsevier Ltd. (b) The X-ray fluorescence mapping of Cr-doped $SrTiO_3$ manifests the fact that the oxygen vacancies are electrochemically formed around the metallic electrode, accumulate, and construct the conducting region within the oxides. Reproduced with permission.[40] Copyright 2009, Wiley-VCH. (c) Hysteretic conductance behavior of the graphene on $SrTiO_3$ thin film (90 nm) device under the systematically expanded gate voltage sweep from -1, +0.5 V to ±4.5 V. Following the alphabetical sequence, this behavior can be understood in terms of generation/annihilation of oxygen vacancies within the $SrTiO_3$ thin film. (d) On top of hysteresis, the quantun Hall conductance of graphene on $SrTiO_3$ thin film appears in the magnetotransport measurement. (e) The mapping of the relationship between dielectric constant, the thickness of oxygen vacancy layer ($\varepsilon_{OV}$, $d_{OV}$) as well as the range of gate field. Reproduced with permission.[41] Copyright 2017, Wiley-VCH.

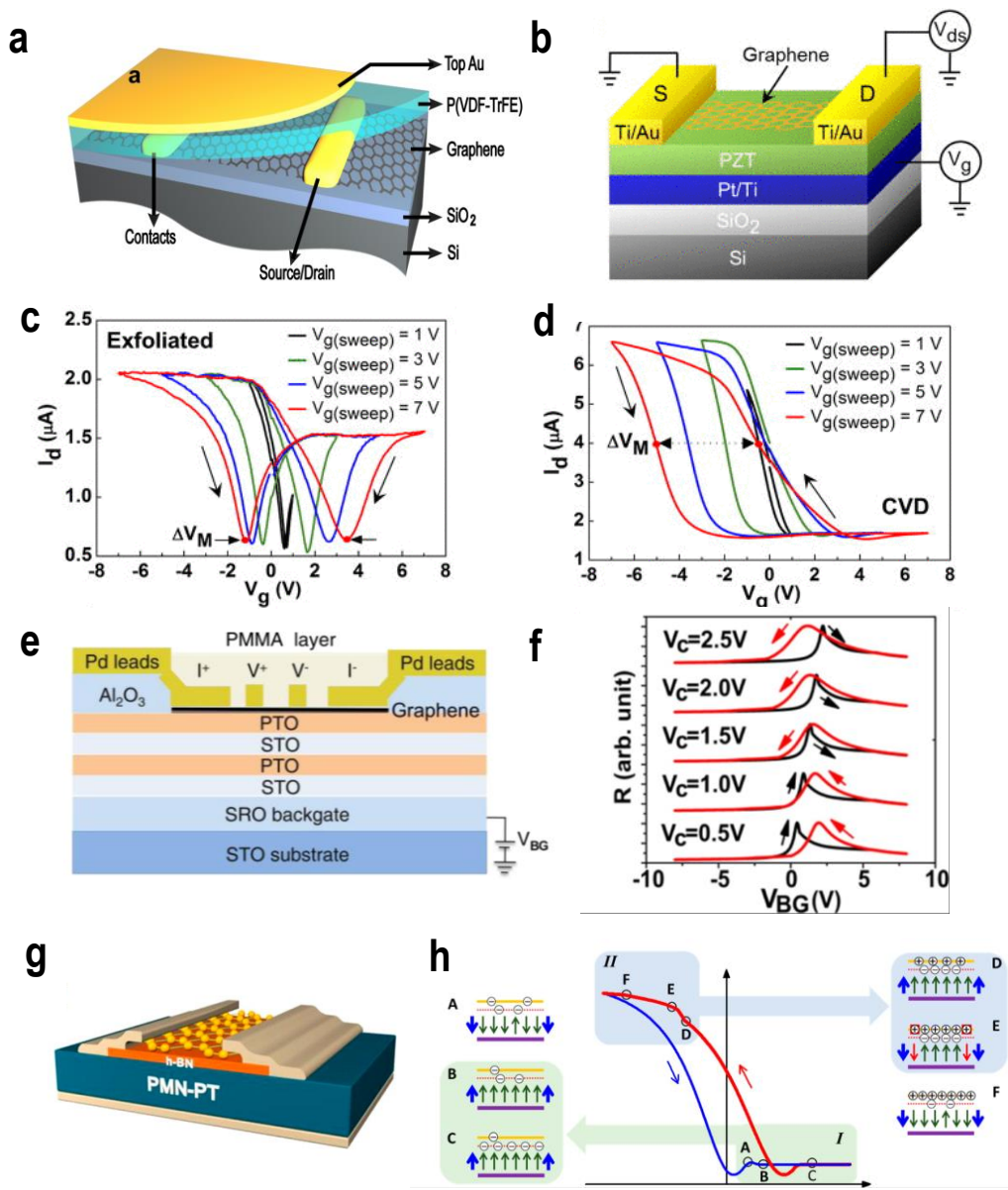

**Figure 5.**
(a) Graphene/ferroelectric-polymer poly(vinylidene fluoride-trifluoroethylene) field-effect transistor (FET): device schematic. Reproduced with permission.[55] Copyright 2009, American Institute of Physics. (b-d) Graphene on thin-film ferroelectric-oxide Pb(Zr,Ti)$O_3$ in the back-gate FET configuration: device schematic ((b)), the hysteretic $I_{Drain}$-$V_{Gate}$ curves of exfoliated graphene ((c)) and those of CVD-synthesized graphene ((d)). Reproduced with permission.[62] Copyright 2011, American Institute of Physics. (e-f) Graphene on ferroelectric superlattice PbTi$O_3$-SrTi$O_3$ substrate: device schematic ((e)) and simulated $I_{Drain}$-$V_{Gate}$ curves ((f)) under the assumption of a fast-trapping process, where hysteresis direction changes as a function of gate-voltage sweep range. Reproduced with permission.[70] Copyright 2014, American Chemical Society. (g-h) Graphene on ferroelectric single-crystalline [Pb(Mb$_{1/2}$Nb$_{2/3}$)$O_3$]$_{1-x}$-[PbTi$O_3$]$_x$ (PMN-PT) substrate where hexagonal BN flake is inserted as an interfacial layer between graphene and PMN-PT: device schematic ((g)) and the model $I_{Drain}$-$V_{Gate}$ curves explaining the conductance saturation in the region I and the conductance jump in the region II ((h)). Reproduced with permission.[73] Copyright 2015, American Chemical Society.

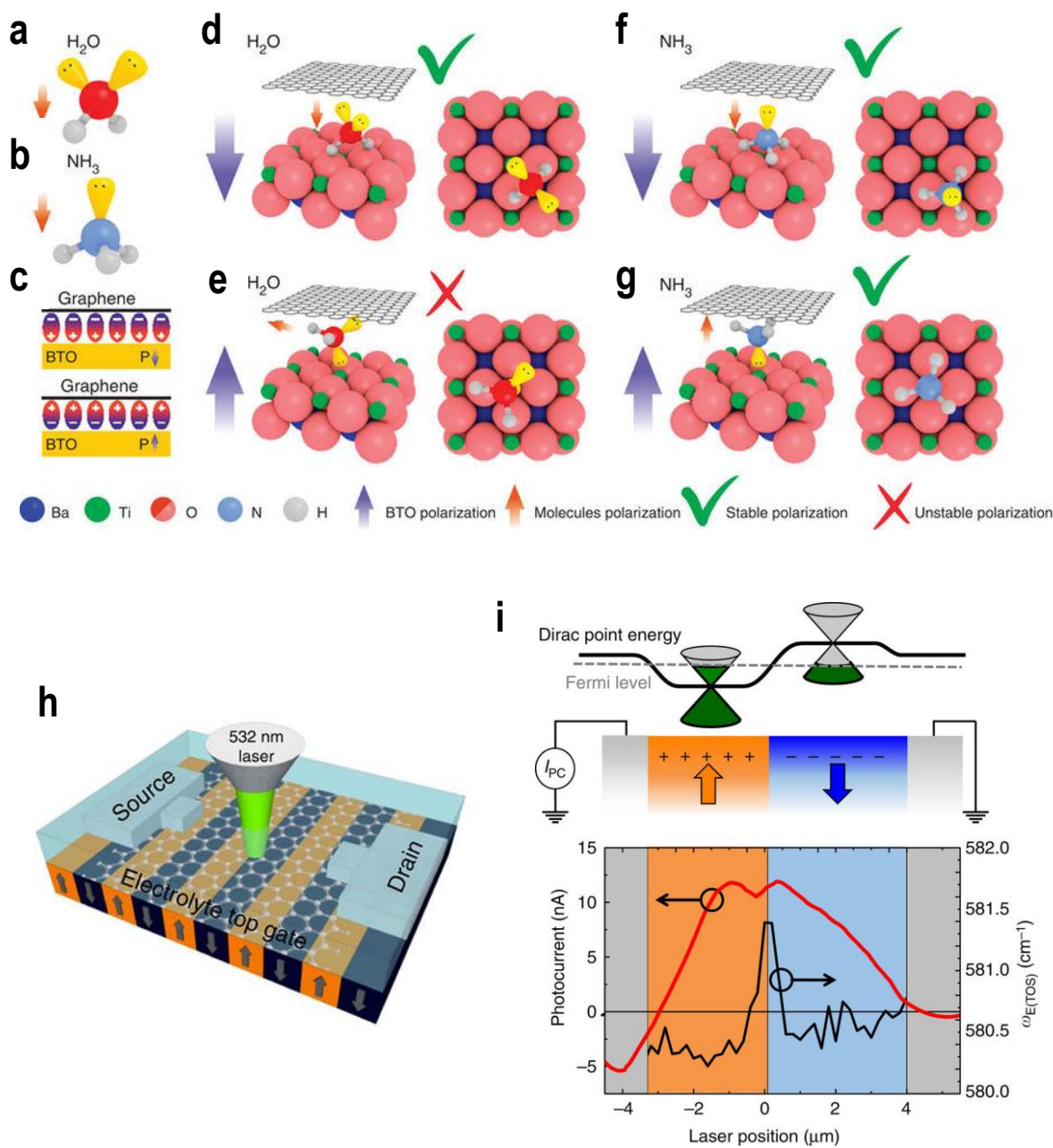

**Figure 6.**
(a-g) Graphene/ferroelectric-oxide BaTiO$_3$ (BTO) vertical tunneling devices: pictures of polar molecule water ((a)) and ammonia ((b)); possible orientations of molecular dipoles at the graphene/BTO interface translating BTO's ferroelectric polarization direction ((c)); illustrations showing water molecule between graphene and BTO with downward or upward polarization ((d-e)); illustrations same as (d) and (e) for ammonia molecule ((f-g)). Reproduced with permission.[77] Copyright 2014, Nature Publishing Group. (h-i) Schematic of graphene on periodically poled LiNbO$_3$ device ((h)) and its photocurrent response due to the ferroelectricity-induced p–n junction in graphene ((i)). Reproduced with permission.[81] Copyright 2015, Nature Publishing Group.

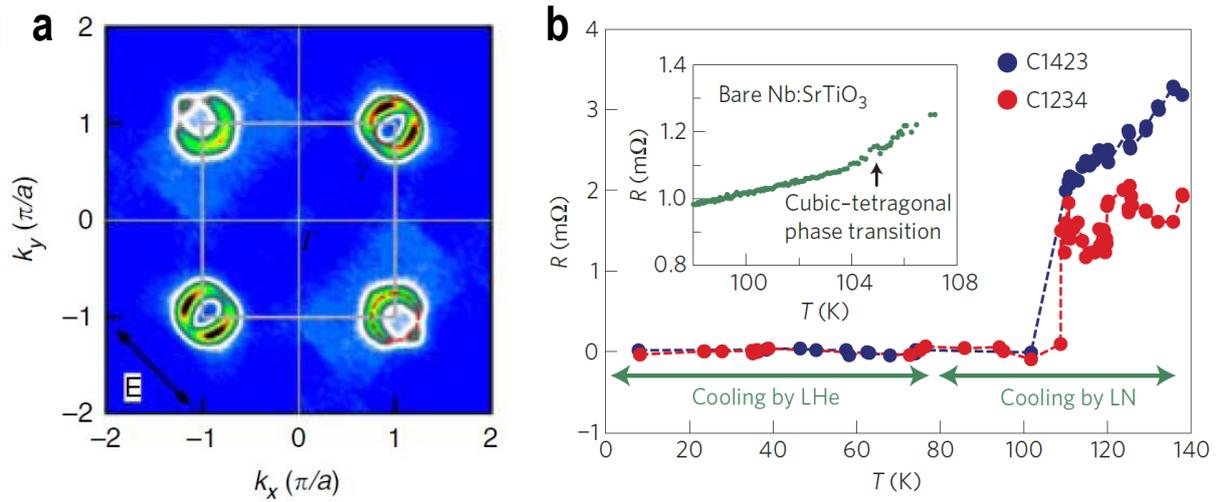

**Figure 7.**
(a) Fermi surface of the single-layer FeSe on SrTiO$_3$ at 20 K, showing only the electron-like Fermi surface sheet around $M(\pi, \pi)$. Reproduced with permission.[87] Copyright 2012, Nature Publishing Group. (b) Temperature dependence of the resistance of the single-layer FeSe film on SrTiO$_3$, obtained from a linear fit to the *I–V* curves measured by the in-situ transport measurement. The inset shows the temperature dependence of resistance taken on a bare SrTiO$_3$ surface. Reproduced with permission.[82] Copyright 2014, Nature Publishing Group.

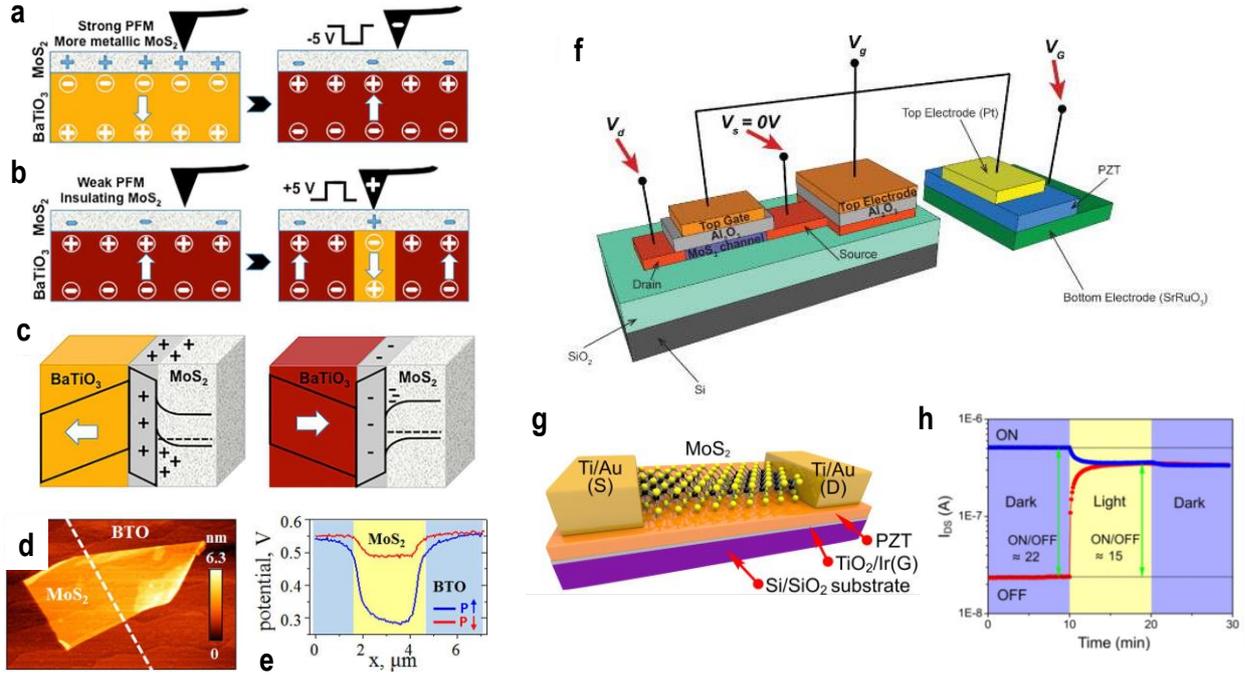

**Figure 8.**
(a-e) Asymmetric switching behavior of the MoS$_2$–BaTiO$_3$–SrRuO$_3$ tunnel junction: schematics of complete switching from the downward to the upward state by a negative voltage pulse ((a)) and partial switching from the upward to the downward state by a positive voltage pulse ((b)); simplified picture of polarization-induced band structure change ((c)); topographic image of the MoS$_2$ flake on the surface of the 12 unit-cell-thick BaTiO$_3$ film ((d)); surface potential profiles along the dashed line marked in (d) depending on the polarization directions ((e)). Reproduced with permission.[108] Copyright 2017, American Chemical Society. (f) Schematic of the MoS$_2$ channel transistor whose gate is connected to an external Pb(Zr,Ti)O$_3$ capacitor. Reproduced with permission.[110] Copyright 2018, American Institute of Physics. (g-h) MoS$_2$-Pb(Zr,Ti)O$_3$ ferroelectric FET: device schematic ((g)) and effect of visible light illumination on the data retention characteristics. Reproduced with permission.[112] Copyright 2015, American Chemical Society.